\documentclass[aps,pra,showpacs,twocolumn]{revtex4}
\usepackage{bm}
\usepackage{amsmath}
\usepackage{amssymb}
\usepackage{graphicx}
\usepackage{epsfig}
\usepackage{epstopdf}
\begin{document}

\def\pr{\prime}
\def\be{\begin{equation}}
\def\en#1{\label{#1}\end{equation}}
\def\d{\dagger}
\def\bar#1{\overline #1}
\newcommand{\per}{\mathrm{per}}

\newcommand{\rd}{\mathrm{d}}
\newcommand{\vare}{\varepsilon }

\title{     Tight bound on trace distance  between  a realistic   device with partially indistinguishable bosons and the ideal Boson Sampling    }

\author{V. S. Shchesnovich}

\address{Centro de Ci\^encias Naturais e Humanas, Universidade Federal do
ABC, Santo Andr\'e,  SP, 09210-170 Brazil }

\begin{abstract}
We study  the  closeness of an experimental unitary bosonic network with only partially indistinguishable bosons in an arbitrary mixed input state, in particular an  experimental realization of the Boson Sampling,  to the ideal  bosonic network, where the measure of  closeness of two networks is the trace distance between the output probability distributions.    An upper bound on the trace distance to the ideal bosonic network  is proven and also a bound on the difference between  probabilities of an output configuration.  Moreover, the  upper bound on the trace distance is tight, provided  that a  physically transparent distinguishability conjecture is true.   For a small distinguishability error it is shown that    a realistic device with $N$ bosons  is at a  constant trace distance to the ideal Boson Sampling    under  the  $O(N^{-1})$-scaling of  the mismatch of internal states of bosons.
 \end{abstract}

\pacs{ 42.50.St, 03.67.Ac, 42.50.Ar }
\maketitle

\section{Introduction}
\label{sec1}

The Boson Sampling (BS)   of   S.~Aaronson and A.~Arkhipov \cite{AA}  can potentially become  the first operating  device with  postclassical computational power, if scaled up to few dozens of bosons.  It can be implemented with  passive linear  optical    devices,    single-photon sources, and non-adaptive detection of photons, clearly    not enough for   the universal quantum computation  \cite{KLM,BarSan,BookNC}. A complex  interference of  bosonic  path amplitudes in a unitary linear network elevates the  output  probability distribution  of the BS    to the $\#$P class of computational   complexity \cite{Valiant,C,ALOP}, asymptotically inaccessible for a classical computer even to verify the result.       The key point in optical implementation of the BS   is  the quantum   indistinguishability of   single photons,  resulting in    the Hong-Ou-Mandel  dip  \cite{HOM} (see also Ref. \cite{LB}).    Several  labs have  performed the proof of principle experiments on small networks with few single  photons    \cite{E1,E2,E3,E4,E5}.   Various   error models  have been   analyzed \cite{R1,R2,R3,NDBS}. It is now clear      that an experimental  device with postclassical computational power is feasible, if   setup errors are properly tuned \cite{BSscal}.    An experimental implementation can now be chosen between various   setups   \cite{BSIons,GBS,TBS}.    Though  no  unconditional verification  method  of an operating device is known, there are  results in this direction  \cite{notUniform}   with an experimental confirmation \cite{ExpValid,ExpComplLinOpt}. Moreover,  there is a  conditional  certification protocol based on many-body interference    \cite{ZeroProb}. Recently, it was shown that the Franck-Condon transition probabilities in molecular vibronic spectra can be simulated on the BS  with a modified input \cite{MVS}.

In the present work we consider how   to measure the    closeness between  a realistic BS device, implemented  with only partially indistinguishable bosons, and the ideal BS. We use the trace distance   between the output probability distributions of a  realistic   device and  the ideal  BS  as the measure of their closeness and study how to estimate   the trace distance    using the information derivable from  experiment.    We give an upper   bound on the  trace distance, and show that under a highly plausible physically clear conjecture the derived bound  is tight.  We employ a recently developed partial indistinguishability theory \cite{PartIndist}  in our derivation of the upper bound on the trace distance. In fact, we derive the main results  in a  more general form, so that they can be applied to measure   the closeness of an arbitrary  experimental bosonic network, with an arbitrary (mixed) input state of bosons,  to the ideal bosonic one with the same  distribution of  particles over the input modes. Thus one can use the upper bound derived in the present work to quantify the ``bosonness'' of an experimental network. Moreover,      the upper bound on   the trace distance  between a realistic and the ideal bosonic networks involves     the probability of the maximally  bunched output (when all particles are  in the same output mode), which can be estimated  from the  experimental data.

Here it should be noted that the purity   measure of partial  indistinguishability,  proposed  in Ref.~\cite{PartIndist},   which quantifies the quality of a quantum interference in an unitary linear   network (for instance, it reduces to the   Mandel's parameter    \cite{Mandel1991} for two bosons)  is not suitable, in general,  for attesting how close is a  realistic   network  to the ideal \textit{bosonic} network  (i.e., with the output amplitudes  proportional to  the matrix permanents of submatrices of the network matrix).  This is due to the fact that both   species of  identical particles, bosons and fermions, can behave in  an interference experiment as  the other species, if the particles are  in an  appropriate (entangled) input state  (one can  even  simulate the BS  with fermions)  \cite{BFduality}.  For instance,   the output probability distribution  of a bosonic network with bosons maximizing the purity  measure of Ref.~\cite{PartIndist} (thus being  completely indistinguishable particles) can be identically equivalent to that of the ideal  fermionic one, if the input state is  anti-symmetric with respect to permutation of  internal states of  particles.  In the latter case the bosonic network is still a perfectly quantum network (that is why it has the  maximal  purity), but equivalent  to the ideal fermionic network. Only in the case when  the  input state of bosons is just a small perturbation of a state symmetric with respect to internal states of particles,   the purity measure can be used  to quantify how close such a network is to the ideal bosonic one.  In general case,  to measure the closeness to the ideal bosonic network  a new   specific measure  is required, which must reflect  how the  internal state of bosons (the projection of the input state of particles on the Hilbert space of the degrees of freedom   not affected by a network) is close  to  a nearest    symmetric    state, since only such  a symmetric input state  results   in the   ideal bosonic behavior  \cite{PartIndist}.

Since there is  a continuous family of input states  for every  probability distribution realizable at output of an unitary network \cite{PartIndist},  it is  more convenient to work with  the   partial indistinguishability matrix  ($J$-matrix)  of Refs. \cite{NDBS,PartIndist}, which is the unique    image of all input states of  bosons resulting in   the same output probability distribution in a unitary network.  A linear space  $\mathcal{V}_N$   of dimension $N!$ for $N$ particles can be  associated with    $J$-matrices, where vectors and matrices are labeled by elements (permutations) of the symmetric group $\mathcal{S}_N$.  \textit{A crucial point is that  a boson counting measurement at network output is formally a measurement also in the auxiliary linear space $\mathcal{V}_N$.}  Thanks to this fact, by application of the    standard methods  we  derive a uniform (over all networks)   upper  bound on the trace distance of an output probability distribution  to  that of the  ideal bosonic network, which supersedes   the probabilistic bound of Ref. \cite{NDBS}.  Under a physically transparent distinguishability conjecture, there is also   a \textit{lower} bound on the   trace distance,  which has   the same scaling in the total number of bosons as the upper bound, thus  making the latter  tight.
The distinguishability conjecture is a restricted from of a  highly  plausible property that different $J$-matrices  result in distinguishable probability distributions,  when the latter are compared over all possible networks.

In the case of bosons in pure spectral states, our upper bound on the trace distance   to the ideal bosonic network  is proportional  to  a recently proposed permanental measure \cite{Tichy2014}.  However, our   bound is   tighter by  $1/N!$ (the proportionality factor) for $N$ bosons.

The  rest of the text is organized as follows. In an auxiliary section \ref{sec2A} we give a brief account of the  results of Ref. \cite{PartIndist} used below. In section \ref{sec2B} we prove \textit{Theorem 1} (the main result), giving the upper bound on the trace distance to the output probability distribution of the ideal bosonic network. Some mathematical details of the derivation are placed in appendix \ref{appA}. In section \ref{sec2C} we consider a special case of bosons in pure internal states.  In section \ref{sec2D} we derive a bound on the difference between  probabilities of an output configuration, \textit{Theorem 2} (with some mathematical details placed in appendix \ref{appB}).  In section \ref{sec3} we study the limit of a small distinguishability error  and give a scaling law of  the mismatch of  internal states of bosons   with the total number of particles  for a constant trace distance to the ideal bosonic network. In section \ref{sec4} we consider a lower bound on the trace distance and argue that under a highly plausible distinguishability conjecture our upper  bound is tight. In section \ref{sec5} we give concluding remarks.

\section{Main results}
\label{sec2}

\subsection{Output probability formula for a general case of partially indistinguishable bosons}
\label{sec2A}

Let us first recall the necessary details of the partial indistinguishability theory  in a unitary linear bosonic network (see for more details    Ref. \cite{PartIndist}).

Consider a linear  unitary $M$-mode   network   with $N$ identical  bosons  at its input, such that each input mode  $k=1,\ldots,M$ receives a certain number of particles $n_k\ge0$.  The single-particle  degrees of freedom are divided into two parts. One part  is composed of modes  operated on by a linear unitary  network. The other part  consists of internal states   unaffected by the network (whose  Hilbert space will be denoted by $\mathcal{H}$).  A  network is  given by a  unitary matrix $U$, which leaves the  Hilbert space of internal states invariant, i.e.,  acting as follows
\be
a^\dag_{k,j} = \sum_{l=1}^M U_{kl} b^\dag_{l,j},
\en{aUb}
where $a^\dag_{k,j}$/$b^\dag_{k,j}$ creates a boson   in an  input/output mode $k$  and  an internal state $|\phi_j\rangle$ (the set of  states $\{|\phi_1\rangle,\ldots,|\phi_j\rangle,\ldots \}$ is a  basis of $\mathcal{H}$).    A general input state of configuration $\vec{n} = (n_1,\ldots,n_M)$  (we will use such vector notations below)  reads $\rho(\vec{n}) = \sum_{i}q_i|\Psi_i(\vec{n})\rangle\langle\Psi_i(\vec{n})|$  with  $q_i\ge 0$, $\sum_i q_i =1$,  and
\be
|\Psi_i(\vec{n}) \rangle = \frac{1}{\sqrt{\mu(\vec{n})}}\sum_{\vec{j}}C^{(i)}_{\vec{j}}\prod_{\alpha=1}^N a^\dag_{k_\alpha,j_\alpha}|0\rangle,
\en{E1}
where $\mu(\vec{n}\,) \equiv \prod_{k=1}^Mn_k!$, $k_1,\ldots,k_N$ are (generally  repeated) input modes, and $|0\rangle$ is the vacuum state.

Due to permutation symmetry of boson creation operators, the expansion coefficients $C^{(i)}_{\vec{j}}$  can be chosen to satisfy  permutation symmetry with respect to  a  Young  subgroup of the symmetric group $\mathcal{S}_N$, consisting  of permutations  $\pi$ of internal states of bosons   in each  input mode between themselves, $\pi\in \mathcal{S}_{n_1}\otimes \ldots \otimes \mathcal{S}_{n_M}\equiv \mathcal{S}_{\vec{n}}$, where $\mathcal{S}_n$ is the symmetric group of  $n$ objects. Such coefficients are normalized by  $\sum_{\vec{j}} |C^{(i)}_{\vec{j}}|^2 =1$.

For a given unitary network, probability of detection of an output  configuration $\vec{m} = (m_1,\ldots,m_M)$,  $|\vec{m}|=N$ with  $|\vec{m}| \equiv m_1+\ldots +m_M$, depends on the degree of partial indistinguishably of bosons, defined by their internal states.  Moreover, experimental particle detection is   imperfect,  prone to particle losses and  dark counts, besides introducing an effective filtering of the internal states due to  non-ideal sensitivity of detectors. We assume that such effects are small and consider the postselected case of all bosons being detected at  the output of a network.   There still remains  to consider the effect of non-ideal sensitivity of detectors (e.g., of limited spectral  width in the case of photons). Here we  adopt the  simplest model of identical  detectors. For  identical detectors  the input state  alone  determines the bound on the postselected  probability distribution at output of a unitary network (since in this case the same $J$-matrix, defined below,  is assigned to any  output configuration of bosons).

The particle counting measurement is therefore described by the following positive operator-valued measure   $\{\Pi(\vec{m}),|\vec{m}|=N\}$, where each element reads (see  the derivation for photons   in appendix A of Ref. \cite{NDBS})
\be
\Pi(\vec{m})  = \frac{1}{\mu(\vec{m})} \sum_{\vec{j}}\left[\prod_{\alpha=1}^N \Gamma_{j_\alpha}\right]\prod_{\alpha=1}^Nb^\dag_{l_\alpha,j_\alpha}|0\rangle\langle0| \prod_{\alpha=1}^N b_{l_\alpha,j_\alpha},
\en{E2}
here $\vec{l} = (l_1,\ldots,l_N)$ is a set of output modes, generally repeated,  corresponding to mode occupations $\vec{m}$, $0\le \Gamma_{j}\le 1 $ is sensitivity of a detector to an internal state $|\phi_j\rangle$.  In   the case of  non-ideal photon counting detectors, the summation over discrete internal states is replaced by an integral over spectral states   \cite{PartIndist}.  Probability  of detection of  an output configuration $\vec{m}$ reads
\be
\hat{p}(\vec{m}|\vec{n}) = \mathrm{Tr}\{ \Pi(\vec{m}) \rho(\vec{n}) \}.
\en{E3}
Inserting Eqs. (\ref{E1}) and (\ref{E2})  into Eq. (\ref{E3}) we obtain output probability as a quadratic form, which can be  expressed as  a double sum over $\mathcal{S}_N$~\cite{NDBS,PartIndist}
\be
\hat{p}(\vec{m}|\vec{n}) = \frac{1}{\mu(\vec{m})\mu(\vec{n})} \sum_{\sigma_1,\sigma_2}J_{\sigma_1,\sigma_2}
  \prod_{\alpha=1}^N U^*_{k_{\sigma_1(\alpha)},l_\alpha}U_{k_{\sigma_2(\alpha)},l_\alpha}.
\en{E4}
The partial indistinguishability matrix   $J$  indexed by two  permutations $\sigma_1,\sigma_2\in \mathcal{S}_N$ (which has only up to  $N!$ different elements for identical detectors)  is given by a    trace
 \be
J_{\sigma_1,\sigma_2}  =   \mathrm{Tr}\left(  \hat{\Gamma}^{\otimes N}\rho^{(int)} P_{\sigma_1\sigma^{-1}_2}\right),
\en{E5}
where  $ \hat{\Gamma} \equiv \sum_j{\Gamma}_{j}|\phi_j\rangle\langle\phi_j|$,  $P_\sigma$ is  a representation of $\sigma\in \mathcal{S}_N$
 in  $\mathcal{H}^{\otimes N}$, defined as
\be
P_\sigma |\phi_{j_1}\rangle\otimes \ldots \otimes |\phi_{j_N}\rangle =  |\phi_{j_{\sigma^{-1}(1)}}\rangle\otimes \ldots \otimes |\phi_{j_{\sigma^{-1}(N)}}\rangle,\qquad
\en{E6}
and $\rho^{(int)}$ is an   internal state of bosons corresponding to an input state of Eq.~(\ref{E1}), i.e.,   $\rho^{(int)}  = \sum_i q_i |\Psi_i^{(int)}\rangle\langle\Psi_i^{(int)}|$ with
\be
|\Psi_i^{(int)}\rangle \equiv \sum_{\vec{j}} C^{(i)}_{\vec{j}} \prod_{\alpha=1}^N{\!}^{\otimes} |\phi_{j_\alpha}\rangle.
\en{E7}

The ideal bosonic behavior  in a   unitary linear network occurs for  $J^{(id)}_{\sigma_1,\sigma_2} = 1$ for all $\sigma_{1,2}$. It is known that  this case realizes if  and only if  the internal  state of bosons  satisfies  $S_N \rho^{(int)}S_N =  \rho^{(int)}$ \cite{PartIndist}, where $S_N = \frac{1}{N!} \sum_\sigma P_\sigma$ is the projector on the symmetric subspace of $\mathcal{H}^{\otimes N}$.  In this case     Eq. (\ref{E4}) contains the absolute value squared of the  matrix permanent of a matrix $U[\vec{n}|\vec{m}]$, built from a network matrix $U$   by taking rows with multiplicities  $\vec{n}$ and columns with multiplicities  $\vec{m}$.  For ideal detectors $\Gamma_{l,j}=1$ and
\be
\hat{p}^{(id)}(\vec{m}|\vec{n}) =  \frac{|\mathrm{per}(U[\vec{n}|\vec{m}])|^2}{\mu(\vec{m}) \mu(\vec{n})}.
 \en{p0}
In  particular, the  BS  setup corresponds to  an input with $n_k\le 1$  in the  dilute limit $M\gg N^2$. In this case probability of a bunched output (i.e., with some $m_k\ge 2$) vanishes as $O(N^2/M)$ and in  the asymptotic limit $N\to\infty$ the  output probability distribution   cannot be   simulated   on a  classical computer  \cite{Valiant,ALOP,AA}.   Recently it was argued that for $M \ge N^2$  the   many-body correlations  of indistinguishable particles  may not  vanish  in  the thermodynamic limit \cite{TLim}. On the other hand, it is also known that for $M\ll N$ the output probability distribution can be      approximated to a small multiplicative error by  a semiclassical approach    \cite{Asympt}.

For non-ideal detectors, $\Gamma_{j}\ne 1$, output probabilities $\hat{p}(\vec{m}|\vec{n})$ of Eq. (\ref{E4}) do not sum to 1.  To   compare with the ideal network  Eq. (\ref{p0}),  we consider the postselected  probabilities,  when $N$ input particles are   detected at a network output.   The postselected  output probabilities are given by   division of those of Eq. (\ref{E3}) by the detection probability   $p_d = \sum_{\vec{m}}\hat{p}(\vec{m}|\vec{n})$. We have (see also
 Ref.~\cite{PartIndist})
\be
p_d =   \frac{\mathrm{tr} J}{N!} = \mathrm{Tr}\left( \hat{\Gamma}^{\otimes N}\rho^{(int)} \right),
\en{E9}
where $\mathrm{tr}(A)\equiv \sum_\sigma A_{\sigma,\sigma}$ is the trace in the linear space $\mathcal{V}_N$.

Finally, there is a continuous family of  internal states of bosons resulting in the same   probability distribution at  output of a unitary linear network \cite{PartIndist}. However, it is highly plausible that   no two different   $J$-matrices   give coinciding  output   probability distributions over  all unitary  networks~\footnote{This problem  is  not trivial.    The quadratic form of Eq. (\ref{E4}) is evaluated  on  mutually dependent  variables $Z_\sigma  = \prod_{\alpha=1}^N U_{k_{\sigma(\alpha)},l_\alpha}$, which fact could  indicate  that a given output probability distribution does not define a unique $J$-matrix. However, not every positive  Hermitian  matrix  is a $J$-matrix, since it  must be  given by Eq. (\ref{E5}), possibly resulting in uniqueness of $J$-matrix.}.  Our main results, formulated below,  do not change if the above property is  not true in  general (see also section \ref{sec4}). 

\subsection{ Upper bound on the trace distance to the ideal bosonic network}
\label{sec2B}

We   will work in the  linear space $\mathcal{V}_N$, where $J$-matrix lives (for convenience,  Dirac's notations will be used   for vectors in $\mathcal{V}_N$, e.g., a vector of the natural  basis, in which $J$-matrix is given  by   Eq.~(\ref{E5}), will be  denoted by  $|\sigma\rangle$, $\sigma\in \mathcal{S}_N$).   We consider how close is the  postselected output probability distribution  of a realistic bosonic network to the ideal bosonic network. The postselected  output probability distribution  will be denoted by  $\mathbf{p}$, whereas   a particular output probability by $p(\vec{m})$ (dropping the  input configuration argument $\vec{n}$). The output probability distribution of the ideal bosonic network  is denoted by $\mathbf{p}^{(id)}$,  a particular output probability $p^{(id)}(\vec{m})$ is given by  Eq. (\ref{p0}). Our goal is to estimate the trace distance
\be
D(\mathbf{p}^{(id)},\mathbf{p}) = \frac12 \sum_{\vec{m}} \bigl|p^{(id)}(\vec{m}) - p(\vec{m})\bigr|.
\en{E12}
 The trace distance is   a 1-norm based measure of closeness between two probability distributions (all norm based measures  in a finite-dimensional linear space are equivalent). It is used  in discussion of the computational complexity of a non-ideal  BS  (for details, see Ref. \cite{AA}).  The following result is found.

\textit{ Theorem 1.--} The trace distance Eq. (\ref{E12}) satisfies
\be
 D(\mathbf{p}^{(id)},\mathbf{p}) \le   1 - p_s,
\en{E21A}
with $p_s$ defined   as follows
\be
p_s = \frac{1}{p_d}\mathrm{Tr}\left\{  \hat{\Gamma}^{\otimes N} S_N\rho^{(int)}S_N\right\} = \frac{1}{p_d}\mathrm{Tr}\left\{  \hat{\Gamma}^{\otimes N} \rho^{(int)}S_N\right\}
\en{E23}
(the second form is due to  mutual commutation of    $S_N$ and $ \hat{\Gamma}^{\otimes N} $ and that   $S_N$ is a projector).

Note that $p_s$   is the postselected (for non-ideal detectors) probability of a   symmetric internal  state of bosons at a network input. For a symmetric input state $S_N\rho^{(int)} = \rho^{(int)}$ we get $p_s=1$. In section \ref{sec2C} below it is shown that the only parameter in the upper bound, $p_s$, can be estimated from the experimental data, i.e.,  even when no knowledge about  the input state of particles is available.

 \textit{Proof of Theorem 1.--} The bound of Eqs.~(\ref{E21A})-(\ref{E23})  easily  follows from the following observation.  \textit{Boson counting measurement at a network output can be thought of  as a ``measurement'' also  in the auxiliary linear space $\mathcal{V}_N$.}   Indeed, let us consider the trace-normalized $J$-matrix defined as follows  (see Eq. (\ref{E9}))
\be
\mathcal{J}_{\sigma_1,\sigma_2} \equiv \frac{J_{\sigma_1,\sigma_2} }{N!p_d} = \frac{1}{N!p_d}\mathrm{Tr}\{  \hat{\Gamma}^{\otimes N}  \rho^{(int)} P_{\sigma_1\sigma^{-1}_2}\}.
\en{formJ}
This   $\mathcal{J}$-matrix    has all the properties of  a density matrix in  the linear space $\mathcal{V}_N$,  whereas  the  following vectors $|Z_{\vec{l}}\rangle \in \mathcal{V}_N$   (here $\vec{l} \equiv (l_1,\ldots,l_N)$ the output mode indices),   in the standard basis defined by
\be
\langle \sigma |Z_{\vec{l}}\rangle  \equiv \frac{1}{\sqrt{\mu(\vec{n})}}\prod_{\alpha=1}^N U_{k_{\sigma(\alpha)},l_\alpha},
\en{EQ2}
define  a positive operator-valued measure in $\mathcal{V}_N$:
\be
\sum_{\vec{l}} |Z_{\vec{l}}\rangle\langle Z_{\vec{l}}\,| = \mathcal{P},
\en{EQ3}
with the projector $\mathcal{P}$ given as
\be
\mathcal{P}_{\sigma,\sigma^\prime} = \frac{1}{\mu(\vec{n})}\sum_{\pi\in  \mathcal{S}_{\vec{n}}}  \delta_{\sigma^\prime,\pi\sigma},\quad \mathrm{tr}(\mathcal{P}) = \frac{N!}{\mu(\vec{n})}
\en{P}
 (see appendix \ref{appA} for more details). The  projector $\mathcal{P}$  acts as the identity operator on any $\mathcal{J}$-matrix: $ \mathcal{P} \mathcal{J} =  \mathcal{J} \mathcal{P} =  \mathcal{J}$, due to the symmetry   $\mathcal{J}_{\pi\sigma_1,\sigma_2} =  \mathcal{J}_{\sigma_1,\pi\sigma_2} = \mathcal{J}_{\sigma_1,\sigma_2}$ for all $\pi$  from the  Young subgroup $\mathcal{S}_{\vec{n}}=\mathcal{S}_{n_1}\otimes \ldots \otimes \mathcal{S}_{n_M}$, as discussed in section \ref{sec2A}.   The postselected output probability   assumes a     ``measurement"  form in the linear space $\mathcal{V}_N$
\be
p(\vec{m})= \sum_{m(\vec{l}\,) = \vec{m}} \langle Z_{\vec{l}}\,| \mathcal{J} |Z_{\vec{l}}\rangle,
\en{EQ4}
where the summation is over all output mode indices $\vec{l}= (l_1,\ldots,l_N)$  corresponding to an output configuration  $\vec{m}$ (in total  $N!/\mu(\vec{m})$  terms).

The rest is an elementary algebra.   The  independence of the sum  $\sum_{\sigma_1}\mathcal{J}_{\sigma,\sigma_1}$ from $\sigma$ (recall that
$\mathcal{J}_{\sigma_2,\sigma_1}$ depends on the relative permutation  $\sigma_1\sigma_2^{-1}$ only) means that such a $\mathcal{J}$-matrix  has an    eigenstate $|s\rangle$, whose expansion in the standard basis reads  $\langle\sigma|s\rangle= {1}/{\sqrt{N!}}$, for any permutation $\sigma$. The corresponding eigenvalue is precisely the postselected probability of a symmetric input state $p_s$. Indeed, from  Eq. (\ref{formJ})  by comparing with Eq. (\ref{E23})  we get
\be
\langle s|\mathcal{J}|s\rangle  = \frac{1}{N!}\sum_{\sigma_1,\sigma_2} \mathcal{J}_{\sigma_1,\sigma_2} = \sum_{\sigma}\mathcal{J}_{\sigma,I} = p_s,
\en{psagain}
where $I$ is the identity permutation. Therefore,  $\mathcal{J}$-matrix as    a convex combination of two matrices
 \be
 \mathcal{J} = p_s |s\rangle\langle s|  + (1-p_s) \mathcal{J}^{(\perp)}, \quad \mathcal{J}^{(\perp)}|s\rangle = 0.
 \en{J}
 Using Eq.~(\ref{J}) to rewrite the trace distance  (\ref{E12}) as follows $D(\mathbf{p}^{(id)},\mathbf{p}) = (1-p_s)D(\mathbf{p}^{(id)},\mathbf{p}^{(\perp)})$ (with  $\mathbf{p}^{(\perp)}$ corresponding to $\mathcal{J}^{(\perp)}$),  noting that the trace distance is    bounded by $1$,  we get the upper bound of  Eq. (\ref{E21A}). Q.E.D.

Finally,  it is notable that   the upper bound Eq. (\ref{E21A}) coincides with the trace distance between a given $\mathcal{J}$-matrix and the pure $\mathcal{J}^{(id)}$-matrix of the ideal case $\mathcal{J}^{(id)} = |s\rangle\langle s|$. We have~\footnote{Eq. (\ref{trDJ})  follows from the general property \cite{BookNC}
\mbox{$\mathrm{tr}\{|\rho-\sigma|\} = \mathrm{tr}\{(\rho-\sigma)_+\}$} where $(\rho-\sigma)_+$ is the positive part of the difference between two density matrices, in our case  $(1-p_s)|s\rangle\langle s|$.}
\be
\mathcal{D}(|s\rangle\langle s|,\mathcal{J}) \equiv \frac12\mathrm{tr}\Bigl\{\Bigl||s\rangle\langle s|-\mathcal{J}\Bigr|\Bigr\} = 1- p_s,
\en{trDJ}
where we have used Eq. (\ref{J}).

\subsection{Upper bound on the trace-distance for pure internal states of bosons }
\label{sec2C}

It turns out that, for bosons   in pure internal states,  the  probability $p_s$ defined in Eq. (\ref{E23}) is   intimately related  to the permanental measure  of Ref. \cite{Tichy2014}. Indeed, let $N$ bosons have  pure internal states $|\varphi_\alpha\rangle$, $\alpha=1,\ldots,N$. We have from Eq. (\ref{E5})
\be
J_{\sigma_1,\sigma_2} = \prod_{\alpha=1}^N\langle\varphi_{\sigma_1(\alpha)}|\hat{\Gamma}|\varphi_{\sigma_2(\alpha)}\rangle = \prod_{\alpha=1}^N\langle\varphi_{\alpha}|\hat{\Gamma}|\varphi_{\sigma_2\sigma^{-1}_1(\alpha)}\rangle,
\en{E24}
thus
\be
p_d = \prod_{\alpha=1}^N\langle\varphi_{\alpha}|\hat{\Gamma}|\varphi_{\alpha}\rangle
\en{E25}
and
\be
p_s = \frac{1}{p_d\,N!}\sum_\sigma   \prod_{\alpha=1}^N\langle\varphi_{\alpha}|\hat{\Gamma}|\varphi_{\sigma(\alpha)}\rangle = \frac{\mathrm{per}\left(G\right)}{N!},
\en{E26}
where the Gram matrix built from the internal states  reads $G_{\alpha\beta} = \langle\widetilde{\varphi}_{\alpha}|\hat{\Gamma}|\widetilde{\varphi}_{\beta}\rangle$ with $|\widetilde{\varphi}_\alpha\rangle = |{\varphi}_\alpha\rangle/(\langle\varphi_\alpha|\hat{\Gamma}|\varphi_\alpha\rangle)^{\frac12}$.

However,   our   bound   given by Eq. (\ref{E21A})  is tighter by $1/N!$  than   the  previous  bound, $N! - \mathrm{per}\left(G\right)$,  proposed in Ref. \cite{Tichy2014}.

\subsection{Bound on the variation  of  an output probability from that of the ideal network}
\label{sec2D}

Let us recall that  the case of classically indistinguishable particles  corresponds to $\mathcal{J} =\frac{\mu(\vec{n})}{N!}\mathcal{P}$
  (the maximally mixed $\mathcal{J}$-matrix as allowed by the symmetry $ \mathcal{P} \mathcal{J} =  \mathcal{J} \mathcal{P} =  \mathcal{J}$, see also Ref. \cite{PartIndist}). The classical  probability of an output configuration $\vec{m}$  reads
\be
 p^{(cl)}(\vec{m})  =  \frac{\mathrm{per}(|U|^2[\vec{n}|\vec{m}])}{\mu(\vec{m})},
 \en{pcl}
where $|U|^2[\vec{n}|\vec{m}]$ is a matrix constructed from elements $|U_{kl}|^2$ by taking rows with multiplicities $\vec{n}$ and columns with multiplicities $\vec{m}$.  We are interested in how close is  the  postselected probability of an output configuration $\vec{m}$ in a realistic network to that of the ideal bosonic network. The following result is found.

\textit{Theorem 2.--} The  difference between the postselected  probability  of an output configuration $\vec{m}$ in  a realistic bosonic network and  that in  the ideal bosonic one   satisfies
\be
\bigl|p^{(id)}(\vec{m}) - p(\vec{m}) \bigr| \le (1 - p_s)\frac{N!}{\mu(\vec{n})}p^{(cl)}(\vec{m}),
\en{E29A}
where $p_s$ is  given by Eq. (\ref{E23}).

\textit{Proof of Theorem 2.--} The bound (\ref{E29A})  follows from Eqs. (\ref{EQ4})-(\ref{J}) and the following inequality
\be
\biggl|\sum_{m(\vec{l}\,)=\vec{m}}\langle Z_{\vec{l}}\,|\mathcal{A}|Z_{\vec{l}}\rangle \biggr| \le  \sum_{m(\vec{l}\,)=\vec{m}}\langle Z_{\vec{l}}\,|Z_{\vec{l}}\rangle = \frac{N!}{\mu(\vec{n})}p^{(cl)}(\vec{m}),
\en{Bpm}
valid   for any  Hermitian  matrix $\mathcal{A}$ whose  eigenvalues  lie in the interval $[-1,1]$.  Setting $\Delta \mathcal{J} = |s\rangle\langle s| - \mathcal{J}^{(\perp)}$, noticing that  its eigenvalues lie in the interval $[-1,1]$,  using Eq. (\ref{Bpm}), (\ref{EQ4}),  and (\ref{J})  we get
\begin{eqnarray}
\bigl|p^{(id)}(\vec{m}) - p(\vec{m}) \bigr| & = & (1-p_s) \biggl| \sum_{m(\vec{l}\,)=\vec{m}} \langle Z_{\vec{l}}\,|\Delta \mathcal{J} |Z_{\vec{l}}\rangle\biggr|\nonumber\\
& \le  & (1-p_s)\frac{N!}{\mu(\vec{n})}p^{(cl)}(\vec{m}),
\label{Add1}\end{eqnarray}
i.e., the bound of Eq. (\ref{E29A}). Q.E.D.

For bosons in pure internal states, as in section \ref{sec2C}, Eq.~(\ref{E29A}) contains two factors: (i) the complementary of    the probability of   the  ideal bosonic behavior $1-p_s = 1-\mathrm{per}(G)/N!$ (see Eq. (\ref{E26})) and (ii)  the upper bound of  Eq. (\ref{Bpm}) on  output  probability $p(\vec{m})$.
Recently,  a bound similar  to that of Eq. (\ref{E29A}) was   proved for a    real positive    $G$-matrix of  Eq. (\ref{E26}) and  then numerically validated   for the general case of bosons in pure   states \cite{Tichy2014}.      In this respect, \textit{Theorem 2}, besides being  an  extension  to  general mixed  inputs,   gives  a generally tighter bound than that of Ref. \cite{Tichy2014}, due to $\mu(\vec{n})$ in the denominator.

Finally,  as  was previously noted in Ref. \cite{Tichy2014},  the   probability $p_s$ can be read off from the experiment. Indeed, $p_s$ appears in the output probability  (postselected for non-ideal detectors) of a maximally  bunched output configuration (see Refs.  \cite{E5,MTMK}), say $m^{(B)}_l = N$. In this case
\be
p^{(id)}(\vec{m}^{(B)}) = \frac{N!}{\mu(\vec{n})}\prod_{\alpha=1}^N|U_{k_\alpha,l}|^2 = \frac{N!}{\mu(\vec{n})} p^{(cl)}(\vec{m}^{(B)})
\en{E31}
and  we obtain (details in Appendix \ref{appB})
\be
p(\vec{m}^{(B)}) = p_s\frac{N!}{\mu(\vec{n})}p^{(cl)}(\vec{m}^{(B)}),
\en{E32A}
i.e., the same  expression as in Ref. \cite{Tichy2014},  extended  here to    arbitrary (mixed) internal states of bosons.

\section{Small errors: scaling of the distinguishability error with $N$}
\label{sec3}

Let us consider a practically relevant case of small errors, which allows for an analytic solution.  We assume that    detectors are nearly ideal $\delta\hat{\Gamma}\equiv 1- \hat{\Gamma}\ll 1$  and that bosons, originated from independent sources,  have  internal states which  are approximately the same pure state, i.e., $\rho^{(int)} = \prod_{\alpha=1}^N{\!}^{\otimes} \rho_\alpha$, with $\rho_\alpha = |\phi\rangle\langle \phi| - \delta \rho_\alpha$ (note that $\mathrm{Tr}(\delta\rho_\alpha)=0$ and $\langle\phi|\delta\rho_\alpha|\phi\rangle >0$).  Below we give all expressions to the first order in $\delta\hat{\Gamma}$ and $\delta\rho_\alpha$.

Expanding the tensor product inside the trace in Eq. (\ref{E5})  we get
\begin{eqnarray}
\label{E34}
&& \prod_{\alpha=1}^N{\!}^{\otimes}\left[\hat{\Gamma}\rho_\alpha\right] = \left(|\phi\rangle\langle\phi|\right)^{\otimes N} \nonumber\\
&&- \sum_{\alpha=1}^N \left(|\phi\rangle\langle\phi|\right)^{\otimes (\alpha-1)} \otimes R_\alpha\otimes\left(|\phi\rangle\langle\phi|\right)^{\otimes (N - \alpha)}\qquad
\end{eqnarray}
with
\be
R_\alpha  = \delta\rho_\alpha + \delta\hat{\Gamma} |\phi\rangle\langle\phi|.
\en{E35}
Let us denote by $C_1(\sigma)$ the set of fixed points (1-cycles) of a permutation $\sigma$. Then, computing the trace in Eq. (\ref{E5}) for the expression in  Eq. (\ref{E34})  we obtain
\begin{eqnarray}
\label{E36}
J_{\sigma_1,\sigma_2}  &=& 1  - \sum_{\alpha\in C_1(\sigma_2\sigma^{-1}_1)}\mathrm{Tr}(R_\alpha) - \sum_{\alpha\notin C_1(\sigma_2\sigma^{-1}_1)}\langle\phi|R_\alpha|\phi\rangle\nonumber\\
&=& 1 - \sum_{\alpha=1}^N \langle\phi| \delta\hat{\Gamma} |\phi\rangle  - \sum_{\alpha\notin C_1(\sigma_2\sigma^{-1}_1)}(1 - \mathcal{F}_\alpha),
\end{eqnarray}
where we have introduced the fidelity $\mathcal{F}_\alpha$ of a mixed internal state $\rho_\alpha$ as follows $\mathcal{F}_\alpha = \langle\phi|\rho_\alpha|\phi\rangle$.
 Since the trace of the third  term on the r.h.s. of  Eq. (\ref{E36}) is zero (all  $\alpha$ are fixed points for $\sigma_2= \sigma_1$), using  Eq. (\ref{E9})   we obtain
\be
p_d = 1- \sum_{\alpha=1}^N \langle\phi| \delta\hat{\Gamma} |\phi\rangle.
\en{E37}
Therefore,   from the definition $\mathcal{J}= J/(p_dN!)$ we have
\be
\mathcal{J}_{\sigma_1,\sigma_2}  = \frac{1}{N!}\left[ 1 - \frac{1}{p_d}\sum_{\alpha\notin C_1(\sigma_2\sigma^{-1}_1)}(1 - \mathcal{F}_\alpha)\right].
\en{E38}
From  Eqs. (\ref{psagain}) and (\ref{E38}) we obtain
\be
1- p_s = \frac{1}{p_dN!}\sum_{\sigma}\sum_{\alpha\notin C_1(\sigma)}(1 - \mathcal{F}_\alpha).
\en{E39}
Let us  introduce the minimal fidelity $\mathcal{F} = \mathrm{min}(\mathcal{F}_1,\ldots,\mathcal{F}_N)$.  Then the bound on the trace distance Eq. (\ref{E21A}) can be estimated as follows
\be
1-p_s \le  \frac{1 - \mathcal{F}}{p_dN!}\sum_{\sigma}\left[N-|C_1(\sigma)|\right] =  \frac{N-1}{p_d}(1 - \mathcal{F}),
\en{E40}
where $|C_1(\sigma)|$ is the total number of 1-cycles and we have used the  well-known fact that  $\frac{1}{N!}\sum_\sigma |C_1(\sigma)| = 1$ \cite{Stanley}.

It is seen from Eq. (\ref{E40}) that  a realistic device with $N$ bosons  is at a  constant trace distance to the ideal bosonic network   for  the following scaling behavior of the minimal fidelity  $\mathcal{F} = 1 - O(1/N)$. The scaling law derived here   is weaker by $\sqrt{N}$ then that  obtained previously by a different method in Ref. \cite{NDBS}. In the next section we argue that the upper bound of Eq. (\ref{E21A}), resulting in the  above scaling law, is tight.

\section{Lower   bound on the  trace distance  and distinguishability conjecture}
\label{sec4}

In section \ref{sec2A} we have already mentioned the problem of uniqueness of $\mathcal{J}$-matrix for a given class of output probability distributions. Whereas in  all  of our calculations above such a  $\mathcal{J}$-matrix was used as just an auxiliary  concept whose   uniqueness property was never used, now we have come  to a point when  the   uniqueness property  of $\mathcal{J}$-matrix  plays an essential role  (though  in a formulation not as   general  as in section \ref{sec2A}).

By  Eq. (\ref{J})   the difference between postselected output probabilities  can be cast as
 \be
 p^{(id)}(\vec{m}) - p(\vec{m}) = (1-p_s)\left[p^{(id)}(\vec{m}) -p^{(\perp)}(\vec{m}) \right],
 \en{EQ5}
where   $p^{(\perp)}(\vec{m})$ corresponds to a density matrix  $\mathcal{J}^{(\perp)}$   orthogonal to    $\mathcal{J}^{(id)} = |s\rangle\langle s|$ of the ideal bosonic  input. Note that such a $ \mathcal{J}^{(\perp)}$-matrix  results in a nonclassical   probability distribution at a network output.  In particular,    $\mathcal{J}^{(\perp)}_{\sigma_1,\sigma_2} = \mathrm{sgn}(\sigma_2\sigma^{-1}_1)/N!$   describes completely  indistinguishable bosons in an  entangled (antisymmetric) internal state, simulating the output probability distribution of completely indistinguishable fermions, due to the boson-fermion duality \cite{BFduality}.

Now, let us consider  the maximum (taken over all $M\times M$-dimensional networks $U$) of the trace distance between the output probability distribution of an  input  resulting in a  density matrix $\mathcal{J}^{(\perp)} $ and that of  the ideal bosonic network 
\be
{d} = \max_{\forall U}D(\mathbf{p}^{(id)},\mathbf{p}^{(\perp)}).
 \en{Lim}
Then, by Eq. (\ref{EQ5})  the maximal over all  $M\times M$-dimensional networks trace distance becomes (with the dependence on parameters given explicitly)
\be
\max_{\forall U}D(\mathbf{p}^{(id)},\mathbf{p}) = d(\mathcal{J}^{(\perp)},N,M)(1-p_s(\mathcal{J},N)).
\en{LB}
At the next step the physical  distinguishability between two mutually orthogonal $\mathcal{J}$-matrices plays an essential role.  Let us state this property as    the following.

\textit{Special distinguishability  conjecture. --} There is a   constant  $d_0>0$ bounding  the function $ d(\mathcal{J}^{(\perp)},N,M)$ of Eq. (\ref{Lim})  uniformly away from zero, i.e., $d(\mathcal{J}^{(\perp)},N,M)\ge d_0$ for all $M,N\ge 2$ and  all (physically realizable) $\mathcal{J}^{(\perp)}$ such that $\mathcal{J}^{(\perp)}|s\rangle =0$.

In other words,  we conjecture that   for any  $N,M\ge 2$  one can always distinguish bosons in an internal state $\rho^{(int)}$  such that $S_N\rho^{(int)}S_N = 0$ from  the ideal bosonic case $S_N\rho^{(int)}S_N = \rho^{(int)}$ (with  output probability distribution  of Eq. (\ref{p0})) by comparing the respective  output  probability distributions over all unitary networks $U$.

If the above conjecture is true, then the upper bound of Eq. (\ref{E21A}) is tight, i.e., it captures the correct  scaling of the trace distance with $N$ (for instance,  the scaling of  the mutual fidelity  for small errors   derived in  section \ref{sec3} is then also tight). E.g., if we take  $N=2$ the  lower bound  $\mathrm{min}(d(\mathcal{J}^{(\perp)},2,M))$ is  positive , since $\mathcal{J}^{(\perp)}$  is pure and corresponds to  the ideal    fermionic   case (see also Ref.  \cite{BFN2}).

From  a physical point of view,   at least a  restricted version of the above conjecture (to some subset of all possible  $\mathcal{J}^{(\perp)}$-matrices), relevant for a realistic  BS device,  must be  true.  Indeed, it is rather clear that  if  bosons have an internal state  which is  not a symmetric one, but a  perturbation of  a   symmetric state,    then  the  corresponding output distribution  must have a nonzero maximum distance to the ideal bosonic network.

\section{Conclusion}
\label{sec5}

We have employed a recently developed approach to partially indistinguishable bosons in a multi-mode unitary linear network for derivation of the upper and lower bounds on the trace distance of the respective output probability distribution to that of the ideal bosonic network. An important property  was used, that in an auxiliary linear space, where the partial distinguishability matrix of bosons is defined, the boson counting measurement at a network output is  formally represented also  as a measurement. This fact allowed us  to  derive the upper and lower bounds on the trace distance by application of the standard methods.
The form of the upper bound has a very clear physical interpretation: it is equal to  the complementary  of the probability of a symmetric input state of bosons, i.e., a state resulting in the ideal bosonic network.

In the case of a small distinguishability error  there is a scaling law of  the  mismatch of internal states of bosons,  stating that   a realistic device with $N$ bosons  is at a  constant trace distance to the ideal bosonic network, if the minimal fidelity of internal states of bosons satisfies  $\mathcal{F} = 1 - O(1/N)$. This scaling law    surpasses  that  obtained previously by a different method in Ref. \cite{NDBS}.

The derived upper  bound is tight (i.e., the lower bound has the same scaling in the number of particles as the upper one)  provided that  a physically clear   distinguishability conjecture  is true.  The conjecture  generalizes and  puts in a mathematical form the  universally assumed assumption   that  a perturbation of  an ideal input state of bosons  generally results  in a non-ideal bosonic behavior. This touches upon the very important issue of the computational complexity of the BS. Indeed,  if the distinguishability conjecture is   false,  then there are such non-ideal input states    of \textit{only partially indistinguishable} bosons,  which have the same output probability distribution as \textit{ the ideal} BS  with the completely indistinguishable bosons (since the trace distance between the two distributions would be  zero).

Our main result can be also states as follows. We have given a tight experimentally accessible measure   of ``bosonness"  of an unitary linear network  with only partially indistinguishable bosons.  The measure is   given by the  trace of projection  of the internal state of bosons  on the symmetric subspace of internal states, i.e., by the probability of having an  input state symmetric in the internal degrees of freedom (which is also proportional to   the probability of the maximally bunched output, which can be estimated  from experimental data).

\medskip
\section{acknowledgements}
This work was supported by the CNPq (Brazil).


\begin{appendix}
\section{Details of the proof  of Theorem 1  }
\label{appA}

 Let us first  verify the following property of vectors $|Z_{\vec{l}}\rangle$
\be
\sum_{\vec{l}} |Z_{\vec{l}}\rangle\langle Z_{\vec{l}}\,| = \mathcal{P}, \quad  \mathcal{P}_{\sigma_1,\sigma_2} = \frac{1}{\mu(\vec{n})}\sum_{\pi\in\mathcal{S}_{\vec{n}}}\delta_{\sigma_2,\pi\sigma_1},
\en{E13}
and show that $\mathcal{P}$ is a projector.
Indeed,  by unitarity of a network matrix $U$
\begin{eqnarray*}
&& \sum_{\vec{l}} \langle\sigma_1|Z_{\vec{l}}\rangle\langle Z_{\vec{l}}\,| \sigma_2\rangle = \frac{1}{ \mu(\vec{n})}  \prod_{\alpha=1}^N\sum_{l_\alpha=1}^M U_{k_{\sigma_1(\alpha)},l_\alpha}U^*_{k_{\sigma_2(\alpha)},l_\alpha}  \\
&&= \frac{1}{ \mu(\vec{n})} \prod_{\alpha=1}^N \delta_{k_{\sigma_1(\alpha)},k_{\sigma_2(\alpha)}} = \frac{1}{\mu(\vec{n})}\sum_{\pi\in\mathcal{S}_{\vec{n}}}\delta_{\sigma_2,\pi\sigma_1}.
\end{eqnarray*}
Moreover,  we have
\[
\mathcal{P}^2_{\sigma,\sigma^\prime} = \frac{1}{(\mu(\vec{n}))^2}\sum_{\sigma_1}\sum_{\pi_{1,2}\in  \mathcal{S}_{\vec{n}}} \delta_{\sigma_1,\pi_1\sigma} \delta_{\sigma^\prime,\pi_2\sigma_1} =  \mathcal{P}_{\sigma,\sigma^\prime},
\]
and  $\mathrm{tr}(\mathcal{P}) = \frac{N!}{\mu(\vec{n})}$.

\section{On the probability $p_s$  }
\label{appB}

We need to show that  $p_s$ appears in the probability of a maximally  bunched output configuration, say  $\vec{l}_B = (l,\ldots,l)$ (i.e., $m^{(B)}_l = N$).
Recalling that  $\langle \sigma|s\rangle  = 1/\sqrt{N!}$ we get  (in this case there is just one measurement element for $\vec{m}^{(B)}_l$)
\be
|Z_{\vec{l}_B} \rangle = \sqrt{\frac{N!}{ \mu(\vec{n})}}\prod_{\alpha=1}^NU_{k_\alpha,l}|s\rangle.
\en{B1}
Thanks to Eq.~(\ref{B1}),  we have saturation of the upper bound (\ref{Bpm}) for an output probability of a maximally bunched output   in  the ideal bosonic case,
\be
p^{(id)}(\vec{m}^{(B)}) = \frac{N!}{\mu(\vec{n})}\prod_{\alpha=1}^N|U_{k_\alpha,l}|^2 = \frac{N!}{\mu(\vec{n})} p^{(cl)}(\vec{m}^{(B)}),
\en{B2}
whereas the other probability appearing in  Eq.~(\ref{EQ5}) vanishes, $p^{(\perp)}(\vec{m}^{(B)}) = 0$.  By resolving Eq. (\ref{EQ5})  we obtain the result
\be
p(\vec{m}^{(B)}) = p_s\frac{N!}{\mu(\vec{n})}p^{(cl)}(\vec{m}^{(B)}).
\en{B3}

\end{appendix}


\begin{thebibliography}{99}
\bibitem{AA} S. Aaronson and A. Arkhipov,  	arXiv:1011.3245 [quant-ph]; Theory of Computing \textbf{9},  143 (2013).




\bibitem{KLM} E. Knill, R. Laflamme, and   G. J. Milburn,  Nature \textbf{409} (2001) 46.

\bibitem{BarSan}  S. D. Barlett and B. C. Sanders, J. Mod. Opt. \textbf{50}, 2331 (2003).

\bibitem{BookNC} M. A. Nielsen and I. L. Chuang, \textit{Quantum Computation and Quantum Information}, (Cambridge University Press, Cambridge, 2000).



\bibitem{Valiant} L. G. Valiant, Theoretical Coput. Sci., \textbf{8},  189  (1979).

\bibitem{C} E. R. Caianiello, Nuovo Cimento, \textbf{10}, 1634 (1953); \textit{Combinatorics and Renormalization in Quantum Field Theory}, Frontiers in Physics, Lecture Note Series (W. A. Benjamin, Reading, MA, 1973).



\bibitem{ALOP} S. Aaronson, Proc. Roy. Soc. London A, \textbf{467},  3393 (2008).



\bibitem{HOM} C. K. Hong, Z. Y. Ou, and L. Mandel, Phys. Rev. Lett. \textbf{59},   2044 (1987).

\bibitem{LB}  Y. L. Lim and A. Beige, New J. Phys., \textbf{7},  155 (2005).

\bibitem{E1} M. A. Broome \textit{et al},  Science \textbf{339},   794 (2013).

\bibitem{E2} J. B. Spring \textit{et al}, Science, \textbf{339},  798 (2013).

\bibitem{E3}  M. Tillmann \textit{et al},  Nature Photonics,  \textbf{7}, 540 (2013).

\bibitem{E4} A. Crespi \textit{et al},   Nature Photonics, \textbf{7}, 545 (2013).

\bibitem{E5} N. Spagnolo \textit{et al}, Phys. Rev. Lett. \textbf{111}, 130503 (2013).


\bibitem{R1} P. P. Rohde and T. C. Ralph, Phys. Rev. A \textbf{85}, 022332  (2012).

\bibitem{R2} P. P. Rohde, Phys. Rev. A \textbf{86}, 052321 (2012).


\bibitem{R3} A. Leverrier and R. Garc{\'i}a-Patr{\'o}n,  arXiv:1309.4687 [quant-ph].

\bibitem{NDBS} V. S. Shchesnovich,   Phys. Rev. A \textbf{89}, 022333 (2014).


\bibitem{BSscal}  V. S. Shchesnovich, 	arXiv:1403.4459 [quant-ph].


\bibitem{BSIons} C. Shen, Z. Zhang, and L.-M. Duan, Phys. Rev. Lett. \textbf{112}, 050504  (2014).

\bibitem{GBS} A. P. Lund, A. Laing, S. Rahimi-Keshari, T. Rudolph, J. L. O'Brien, and T. C. Ralph,  Phys. Rev. Lett. \textbf{113}, 100502 (2014).

\bibitem{TBS} K. R. Motes, A. Gilchrist, J. P. Dowling, and P. P. Rohde,  Phys. Rev. Lett. \textbf{113}, 120501 (2014).

\bibitem{notUniform} S. Aaronson and A. Arkhipov,  	arXiv:1309.7460 [quant-ph].

\bibitem{ExpValid} N. Spagnolo \textit{et al}, Nat. Photon. \textbf{8}, 615 (2014).

\bibitem{ExpComplLinOpt} J. Carolan  \textit{et al},  Nat. Photon. \textbf{8}, 621 (2014).


\bibitem{ZeroProb} M. C. Tichy,   K. Mayer, A. Buchleitner, and  K. Molmer,  Phys. Rev. Lett. \textbf{113}, 020502 (2014).

\bibitem{MVS} J. Huh, G. G. Guerreschi, B. Peropadre, J. R. McClean, and A. Aspuru-Guzik, 	arXiv:1412.8427 [quant-ph].

\bibitem{PartIndist} V. S. Shchesnovich,  	arXiv:1410.1506 [quant-ph].

\bibitem{Mandel1991} L. Mandel, Opt. Lett. \textbf{16}, 1882 (1991).

 \bibitem{BFduality} V. S. Shchesnovich, 	arXiv:1412.0279 [quant-ph];  	 to appear in  Int. J. Quant. Inform. (2015).









\bibitem{Tichy2014}  M. C. Tichy,  Phys. Rev. A \textbf{91}, 022316  (2015).



\bibitem{TLim} J.-D. Urbina, J. Kuipers, Q. Hummel, and K. Richter,  	arXiv:1409.1558 [quant-ph].

\bibitem{Asympt} V. S. Shchesnovich, Int. J. Quantum Inform.,  \textbf{11}, 1350045 (2013).


\bibitem{MTMK} K. Mayer, M. C. Tichy, F. Mintert, T. Konrad, and A. Buchleitner, Phys. Rev. A \textbf{83}, 062307  (2011).

\bibitem{Stanley}   R. P. Stanley, \textit{Enumerative Combinatorics}, 2nd ed., Vol. 1 (Cambridge University Press, 2011).




\bibitem{BFN2} F. T\"{o}ppel and A. Aiello, Phys. Rev. A \textbf{88}, 012130  (2013).







\end{thebibliography}
\end{document}